\pdfoutput=1

\documentclass[11pt]{article}

\usepackage{acl}

\usepackage{times}
\usepackage{latexsym}

\usepackage[T1]{fontenc}

\usepackage[utf8]{inputenc}

\usepackage{microtype}

\usepackage{inconsolata}

\usepackage{algorithm}
\usepackage{algorithmic}

\usepackage{adjustbox}
\usepackage{booktabs}
\usepackage{amsmath}
\usepackage{amsfonts}       
\usepackage{float}
	
\definecolor{LightCyan}{rgb}{0.88,1,1}
\definecolor{mygray}{gray}{0.9}

\usepackage{color, colortbl}
\title{BrewCLIP: A Bifurcated Representation Learning Framework for Audio-Visual Retrieval}

\author{Zhenyu Lu \\
  Carnegie Mellon University \\
  \texttt{zhenyulu@andrew.cmu.edu} \\\And
  Lakshay Sethi \\
  Carnegie Mellon University \\
  \texttt{lsethi@andrew.cmu.edu} \\}

\begin{document}
\maketitle
\begin{abstract}
Previous methods for audio-image matching generally fall into one of two categories: pipeline models or End-to-End models. Pipeline models first transcribe speech and then encode the resulting text; End-to-End models encode speech directly.
Generally, pipeline models outperform end-to-end models, but the intermediate transcription necessarily discards some potentially useful non-textual information.
In addition to textual information, speech can convey details such as accent, mood, and and emphasis, which should be effectively captured in the encoded representation.
In this paper, we investigate whether non-textual information, which is overlooked by pipeline-based models, can be leveraged to improve speech-image matching performance.
We thoroughly analyze and compare End-to-End models, pipeline models, and our proposed dual-channel model for robust audio-image retrieval on a variety of datasets.
Our approach achieves a substantial performance gain over the previous state-of-the-art by leveraging strong pretrained models, a prompting mechanism and a bifurcated design. Our code will be made available. 
\end{abstract}

\section{Introduction}
Audio-image matching is similar to the popular text-image matching problem, but it is comparatively less studied due to the increased complexity of modeling audio and the lower availability of paired audio-image data. 
Speech contains intricate information in the form of tone, timbre, stress patterns, and contextual cues that can exhibit substantial variation among speakers and languages. However, this complexity also means that speech carries richer information compared to text. 
\begin{figure}[!t]
\begin{center}
\includegraphics[width=0.46\textwidth]{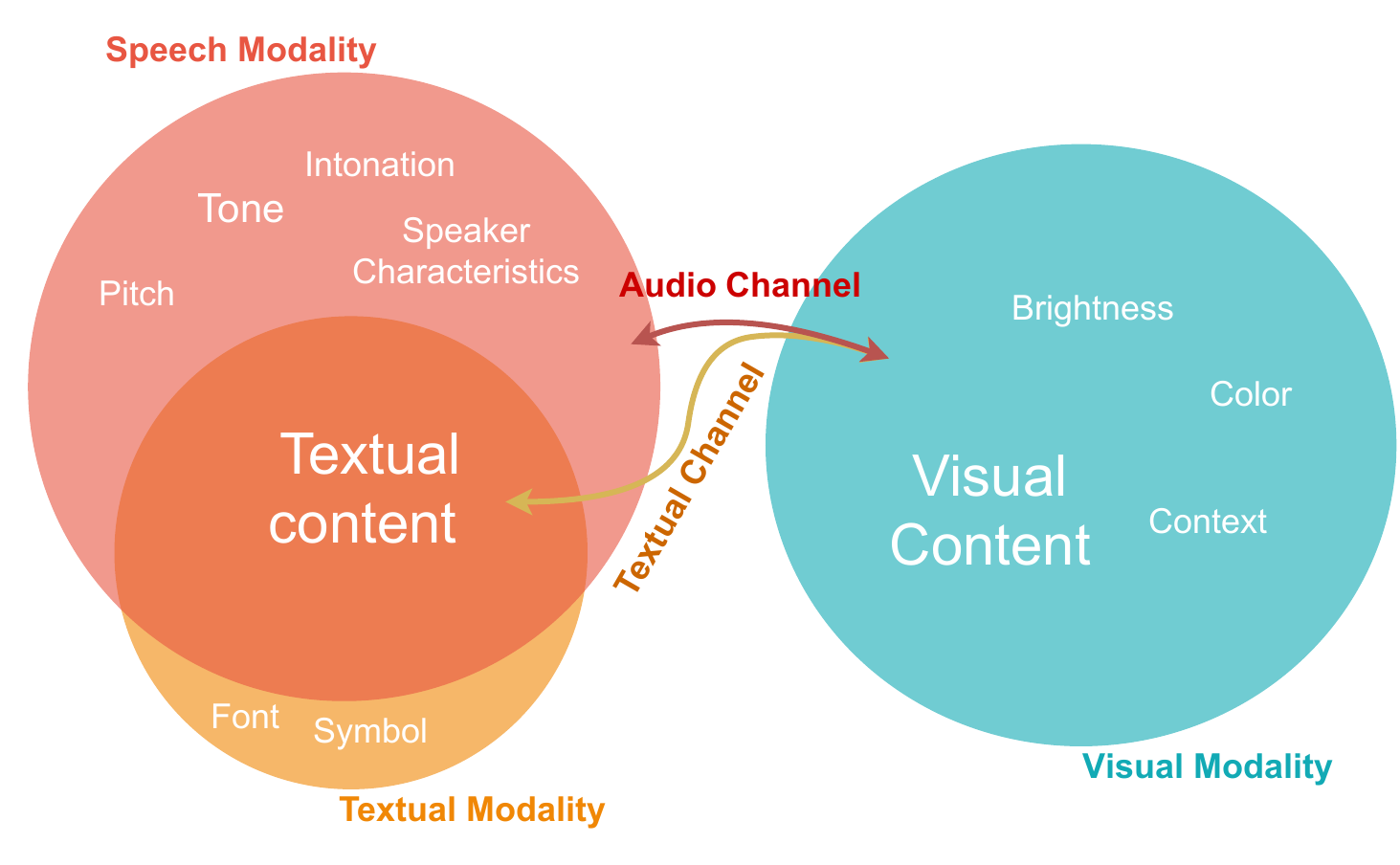}
\vspace{-2ex}
\caption{\textbf{Venn diagram illustrating our dual-channel design.} Our text channel primarily focuses on capturing textual information, while audio channel can complements the textual information and facilitates the communication of non-textual information.}
\label{fig:archecture_simple}
\vspace{-4ex}
\end{center}
\end{figure}


Recently, multi-modal models \cite{CLIP,uniter,vlbert} has proven to be highly beneficial for retrieval tasks and other various downstream tasks. While numerous audio-image models have been proposed \cite{speechclip2022, talkdontwrite2021, Fast-VGS}, none have achieved the remarkable impact seen with recent text-image models. Current efforts in the audio-image matching domain can be broadly categorized into two branches: pipeline models and End-to-End (E2E) models.

Transcription-based pipeline models tend to achieve superior performance due to the availability of well-developed Automatic Speech Recognition (ASR) systems and text encoders. Some studies even consider pipeline models with perfect transcription as the upper bound for performance on the task \cite{Fast-VGS}. However, we argue that although pipeline models generally outperform E2E models, a well-learned audio representation jointly learnt with image should not be limited to textual information alone.



In light of these observations, we raise the question: Can we devise a method to learn a representation that accurately captures textual information while also retaining crucial audio-specific information? Furthermore, does audio contain additional information that is vital for the audio-image retrieval task?
To address these questions, we propose a novel bifurcated model called BrewCLIP. 

BrewCLIP is a modified pipeline model; its base is Whisper\cite{whisper2022} as a transcription model and CLIP\cite{CLIP} as a text and image encoder. 
We augment this frozen pipeline model with an additional End-to-End  audio encoder. 
Comprehensive experiments are conducted to assess the interactive effects of various models (E2E, pipeline, dual-channel), various datasets (scripted and unscripted), and the impact of prompt finetuning on the final performance of the model.
\noindent To summarize, we show:

\begin{itemize}
\item Our proposed model BrewCLIP has surpassed the previous SotA performance for image-text matching.
\item We explore shared prompting as a cheap way to finetune the frozen models, which improves the performance significantly.
\item We probe the final model to explore whether it exploits non-textual emotion information by applying BrewCLIP's End-to-End audio encoder to speech emotion recognition (SER) and show that our model can successfully learn mood information in the utterances.
 \end{itemize} 
\section{Related Work}
\subsection{Vision Language Models}
Recent progress in vision-language models has been made by exploiting powerful jointly-trained unimodal encoders. Text-visual learning has always been the predominant focus, applied to tasks like, image classification or detection \cite{filip,multimodal_pretrain}, text-image retrieval \cite{CLIP}, text conditional image generation\cite{zeroshottext_image,dalle2}, audio-visual learning remains relatively under-explored, especially lack of a unified framework for multiple tasks. Most efforts in this domain have been concentrated at the video level, encompassing tasks like source separation and Localization\cite{tian2021cyclic,tzinis2020into,mo2023av}, temporal synchronization \cite{cooperative,owens2018audio} and audio-video matching \cite{objectsound,zhao2018sound}. 
\subsection{Audio-Image Retrieval Task}
Cross-modal retrieval \cite{multimodal_pretrain, vilt} is considered as one of the gold standard to evaluate the joint representation across different modalities. Likewise, image-speech matching is to associate spoken descriptions with their corresponding images. 
Performing well on image-speech matching requires the model to effectively associate similar information from vastly different audio and image modalities. 
Image-speech matching finds immediate applications in content-based image retrieval, where spoken descriptions are used to retrieve relevant images. Furthermore, we hope our study can pave the road for the development of a strong audio-visual joint representation, which can be used for multiple downstream tasks. 
\subsection{End-to-End model and Pipeline Model}
Existing literature primarily tackles the image-speech retrieval task in one of two ways: End-to-End(E2E) training or ASR based pipeline models. Both approaches have been explored in works such as \cite{talkdontwrite2021}, which provides exemplary implementations of each. End-to-End models are usually composed of a dual-encoder design which directly associate the raw audio and the image. In contrast, pipeline models rely on an ASR model to convert the raw audio into text format, either full transcription or related keywords, transforming the task into a transcription-image retrieval problem. FaST-VGS \cite{Fast-VGS} is an example of an E2E model that incorporates cross-modal attention within its transformer-based structure.
A more recent method, SpeechCLIP\cite{speechclip2022}, employs several large-pretrained models, such as CLIP and Hubert \cite{HuBERTmodel}, to tackle the image-speech retrieval task. It offers two main model variants: parallel and cascaded, each representing explorations of the two aforementioned approaches.
On the other hand, \citet{textualsupervision2020} explores various means of exploiting textual supervision to enhance performance in the speech-image matching task. However, MILAN \cite{talkdontwrite2021} also demonstrate superior performance for non-pipeline models under certain conditions, largely influenced by the performance of the ASR system. 
\begin{figure*}[!htbp]
\vspace{-8ex}
\centering
{\includegraphics[width=15.5cm]{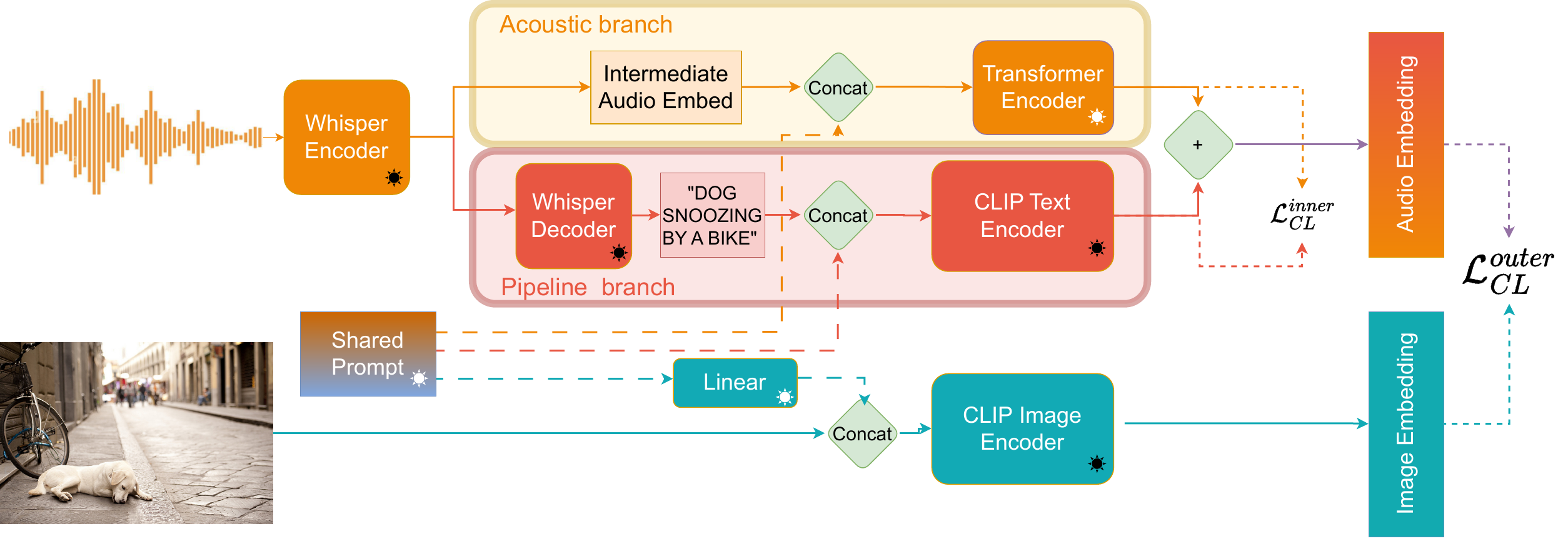}}
\caption{\textbf{Detailed diagram of our proposed model.} The acoustic channel is shaded in yellow and the transcription based pipeline channel is shaded in red. \textbf{E2E-only model} only contains the Acoustic branch and image branch. \textbf{PIP-Only model} only contains the pipeline branch and image branch. \textbf{PIP-only(Zero-shot)} is a special pipeline-based model which is not finetuned by the prompts. \textbf{black Sun: frozen, white Sun: updating.}}
\label{fig:arch_full}
\end{figure*}

\subsection{Prompting}
Prompting has become a prevalent approach for extremely cheap few-shot fine-tuning of transformer-based models. It involves finding (via optimization or heuristic prompt engineering) a set of tokens to prepend to a transformer model's input in order to improve its performance on a downstream task. By keeping the original model's weights frozen, prompted models are highly generalizable. Originally utilized in the NLP domain \cite{parameterefficient,language_fewshot}, prompting has proven useful in the visual domain \cite{vpt,vpt_for_gtl} and has now been adapted for the multi-modal domain \cite{coop,vop} as well.
 
\section{Method}
\subsection{Component Pretrained Models}
\paragraph{\textbf{CLIP}\cite{CLIP}}
CLIP is a widely-used image/text encoder trained on a simple contrastive loss. It has been trained exhaustively on a staggering 400 million web-scraped image-text pairs, yielding a highly generalizable set of encoders. CLIP performs well on images of objects, scenes, and abstract concepts, and its encoders generalize to other vision-language tasks. In light of its strong performance, several variations of CLIP have been proposed. Our approach is inspired by AudioCLIP \cite{audioclip}, Wav2CLIP \cite{wav2clip}, and SpeechCLIP, which apply CLIP-like methods to audio-related tasks. \\
\paragraph{\textbf{Whisper}\cite{whisper2022}}
To leverage CLIP for audio-image matching, we must first transcribe the raw input audio using an Automatic Speech Recognition (ASR) model. Fortunately, current ASR models have reached approximately human-level performance, paving the way for a strong new pipeline-based model. Whisper adopts an encoder-decoder transformer structure, leveraging a large-scale dataset of low-quality paired utterances and their transcriptions from the internet. This approach has enabled Whisper to achieve state-of-the-art performance for speech transcription.


\subsection{Task Definition}


Consider a dataset $D = \{(\boldsymbol{X}_V^{[i]}, \boldsymbol{X}_A^{[i]}) : i \in [n]\}$,
where $(\boldsymbol{X}_V , \boldsymbol{X}_A)$ denotes paired image and audio. Each image or audio clip may appear in multiple pairs. The matching task can be defined as follows: given a query of either modality $\boldsymbol{X}_{V/A}^{[i]}$, the objective is to retrieve an associated instance $\boldsymbol{X}_{A/V}^{[i]}$ from the set of all given data of the other modality. \\
\textbf{Zero-Shot Setting} 
The model utilizes the pre-trained modules without any fine tuning or prompting. It is directly evaluated on the test set of the target dataset without additional training on that specific dataset. \\
\textbf{Self-Supervised Setting} This is typical setting where the model is fine-tuned with all the samples from the target training dataset and then evaluated on the associated test set.

\subsection{BrewCLIP: Bifurcated Whisper CLIP}
Existing studies have revealed a recurring phenomenon where pipeline models tend to outperform End-to-End (E2E) models, particularly when a superior ASR model is available. However, we posit that pipeline models suffer from a crucial drawback: they lose vital information that only exists in the raw audio as they transcribe the audio into text. To investigate this hypothesis and address the issue, we propose a bifurcated design, depicted in Fig. \ref{fig:arch_full}, capable of processing raw audio and its transcription in two distinct but parallel channels. Our architecture builds upon two powerful pre-trained models, Whisper and CLIP, as previously introduced. 
By employing the bifurcated design, we aim to retain essential audio-specific information that might be lost during the transcription process while still benefiting from the strengths of the pre-trained text and image encoders. \\
\textbf{Image Encoder}
We start by tokenizing the raw image $\boldsymbol{X}_V^{[i]}$ into $n$ patches, creating the patch embedding $\boldsymbol{E}_V \in \mathbb{R}^{n \times d_v}$, and these patch embedding $\boldsymbol{E}_V$ are then fed into the pretrained CLIP image encoder $\mathcal{V}$ to obtain the image representation $\boldsymbol{F}_V$.
\[\boldsymbol{F}_V = ImageEncoder(\boldsymbol{E}_V)\]
\textbf{Transcription based Textual Module}
This module follows a typical pipeline approach to obtain the textual embedding from the audio by connecting Whisper and CLIP text encoder.
The Whisper encoder takes the re-sampled audio embedding $\boldsymbol{E}_A$ from the raw $\boldsymbol{X}_A$ and outputs the intermediate audio feature $\boldsymbol{Z}_A$.
\[\boldsymbol{Z}_A = WhisperEncoder(\boldsymbol{E}_A)\]
The decoder then takes the encoded audio features $\boldsymbol{Z}_A$ as input and generate the corresponding transcription. The tokenized transcription embedding $\boldsymbol{E}_T$ is then fed into the text encoder to obtain the Textual feature $\boldsymbol{F}_T$. 
\[\boldsymbol{F}_T = TextEncoder(WhisperDecoder(\boldsymbol{E}_T))\]

\noindent \textbf{Acoustic End-to-End Module}
We augment the baseline pipeline model with an additional End-to-End channel that handles the raw audio features, in order to capture the lost non-textual information. 
Given the intermediate audio embedding $\boldsymbol{Z}_A$ generated from the Whisper encoder, we feed it into a transformer encoder layer to obtain a broad acoustic representation.
\[\boldsymbol{F}_A = TransformerEncoder(\boldsymbol{Z}_A)\]
Subsequently, we fuse $\boldsymbol{F}_A$ and $\boldsymbol{F}_T$ together by simply addition, yielding the final audio representation.
\[\boldsymbol{F}^{Final}_L = \boldsymbol{F}_A + \boldsymbol{F}_T\]
\textbf{Shared Prompting}
Cross-modal interaction has been proved to highly beneficial in models that process multiply modalities. One popular design involves cross-modal attention\cite{multimodal,decoupling}, encouraging the overlap between different modalities to promote learning of shared information. However, when applied to retrieval tasks, cross-modal attention introduces additional challenges\cite{vilt}.
Besides, cross-modal transformers may not be applicable for large pre-trained models. Therefore, in our approach, we finetune the pre-trained model by adopting a lightweight shared prompt design to encourage the commutation between data from different modalities.
We formulate our prompted textual embedding as:
\begin{multline*}
\boldsymbol{\hat E}_T = [\boldsymbol{P}_T^{[1]}, \boldsymbol{P}_T^{[2]},\cdots,
\boldsymbol{P}_T^{[M]}, \\
\boldsymbol{E}_T^{[1]}, \boldsymbol{E}_T^{[2]}, \cdots, \boldsymbol{E}_T^{[K]}] 
\end{multline*}
Where $\boldsymbol{P}_T^{[m]} : (m \in {1,\cdots, M})$ is a vector with
the same dimension as the embeddings, $M$ is a hyperparameter specifying the length of prompts, and $K$ is the length of original textual embedding $\boldsymbol{E}_T$\\
The same $\boldsymbol{P}_T$ is used in the E2E channel by similarly concatenating with the raw $\boldsymbol{Z}_A$, obtaining $\boldsymbol{\hat Z}_A$ .  We connect the two prompts from different modalities using a linear layer, primarily to handle the dimension mismatch issues while also
allowing some flexibility to each individual prompt.
\[\boldsymbol{P}_V = Linear(\boldsymbol{P}_T)\]

\noindent \textbf{Loss Function} We follow the same InfoNCE loss \cite{infonce}  used in CLIP. $\mathcal{L}_{V\rightarrow L} $ is denoted as shown in Equation \ref{eqn:contrastive_loss}).
\vspace{-3ex}

\begin{equation}
\label{eqn:contrastive_loss}
\begin{gathered}
\mathcal{L}_{V\rightarrow L}=\frac{-1}{N}\sum_{i=1}^{N}\log{
\frac{
\displaystyle \exp{\left(\frac{sim(\boldsymbol{F}^{[i]}_V,\boldsymbol{F}^{[i]}_L)}{\tau}\right)}}{
\displaystyle \sum_{j=1}^{N}
\exp{\left(\frac{sim(\boldsymbol{F}^{[i]}_V,\boldsymbol{F}^{[j]}_L)}{\tau}\right)}}}
\end{gathered}
\end{equation}
where the similarity score is computed by dot product between $\boldsymbol{F}_V$ and $\boldsymbol{F}_L$, and $\tau$ is a learnable temperature variable. 
The loss for the audio branch $\mathcal{L}_{L\rightarrow V}$ exactly follows the same design. \\
We adopt an inner-outer loss design. $\mathcal{L}_{CL}^{outer}$ is the main loss function to link the audio and image modalities.  $\mathcal{L}_{CL}^{inner}$ is a helper loss function to simultaneously regulate the learned acoustic embedding, ensuring it is not too distant from its textual counterpart, while facilitating the acquisition of new non-textual information. 
\begin{equation}
\label{eqn:final}
\begin{gathered}
\mathcal{L}_{CL}^{outer}= \mathcal{L}_{V\rightarrow L} + \mathcal{L}_{L\rightarrow V} \\
\mathcal{L}_{CL}^{inner}= \mathcal{L}_{A\rightarrow T} + \mathcal{L}_{T\rightarrow A} \\
\mathcal{L}^{Final}=  \alpha \mathcal{L}_{CL}^{inner} + (1-\alpha) \mathcal{L}_{CL}^{outer}
\end{gathered}
\end{equation}
Where $\mathcal{L}_{A\rightarrow T}$ computes the contrastive loss from $F_A$ to $F_T$, $\mathcal{L}_{T\rightarrow A}$ is computed in the reverse direction and $\alpha$ is a small hyper-parameter to control the contribution of the inner and outer losses.

\section{Experiment}
For the retrieval task, we conduct experiments on both spontaneous and non-spontaneous spoken caption datasets: Flickr8k, SpokenCOCO, and Localized Narratives-COCO, with average utterance lengths of 10.9, 10, and 36.5 words, respectively.
Our hypothesis is that unscripted speech would contain more audio-specific features compared to speech read from text captions. Unscripted utterances often involve disfluencies and filler words, which pose additional challenges for audio processing. Additionally, we seek to explore whether our model could successfully extract non-textual information, primarily mood information, from the audio. To investigate this, we conduct a brief Speech Emotion Recognition (SER) experiment using our well-trained audio encoder on the RADVESS dataset \cite{ravdess}. \\
\textbf{Evaluation metrics} 
Consistent with the literature, we measure the performance of the retrieval task using top-K recall (where K is 1/5/10), which refers to whether the model retrieved the correct element as at least one of its first K guesses. We also vary whether the query is an image or an audio clip, giving us image2speech and speech2image metrics, respectively. 
Furthermore, if there are multiple correct answers (as is the case with images that receive multiple captions), any correct answer in the top K counts as a valid recall.
\subsection{Dataset}
\textbf{Flickr8k}
The baseline dataset we use is Flickr8k\cite{Flickr8k}, which contains 5 captions per image and around 8k images. Flickr8k contains common images and brief, relevant textual captions. After these textual captions were written, the dataset was extended by asking people to read each textual caption. \\
\textbf{SpokenCOCO and LN-COCO}
SpokenCOCO and Localized Narratives (LN-COCO) are both derived from the MS-COCO image dataset \cite{MS_COCO}, which consists of 123k images. SpokenCOCO utilizes the newer Karpathy train-val split and pairs each image with five spoken captions, where viewers are provided with the corresponding text, similar to Flickr8k.
On the other hand, Localized Narratives contains unscripted spoken English descriptions of images that are generally more spontaneous in nature. 
\\
\\
\begin{table*}[htb!]
\begin{adjustbox}{max width=1\textwidth,center}
\begin{tabular}{*{24}{c}}
\toprule                  
                            & \multicolumn{12}{c}{Non-Spontaneous} 
                            & \multicolumn{6}{c}{Spontaneous}\\
                            \cmidrule(lr){3-12}\cmidrule(lr){15-18}
                            & \multicolumn{6}{c}{Flickr8k} 
                            & \multicolumn{6}{c}{SpokenCOCO}
                            & \multicolumn{6}{c}{LN-COCO}
                            \\
                            \cmidrule(lr){3-6}\cmidrule(lr){9-12}\cmidrule(lr){15-18}
                            & \multicolumn{3}{c}{Speech2Image}
                            & \multicolumn{3}{c}{Image2Speech}
                            & \multicolumn{3}{c}{Speech2Image}
                            & \multicolumn{3}{c}{Image2Speech}
                            & \multicolumn{3}{c}{Speech2Image}
                            & \multicolumn{3}{c}{Image2Speech}
                            \\
\cmidrule(lr){2-4}\cmidrule(lr){5-7}\cmidrule(lr){8-10}\cmidrule(lr){11-13}\cmidrule(lr){14-16}\cmidrule(lr){17-19}
Method  & R1  & R5  & R10 & R1  & R5  & R10 & R1  & R5  & R10 & R1  & R5  & R10 & R1  & R5  & R10 & R1  & R5  & R10\\
\midrule
\textbf{Previous Approaches } \\
\midrule
SpeechCLIP-P \cite{speechclip2022}
                    & 39.1 & 72.0 & 83.0 & 54.5 & 84.5 & 93.2 & 35.8 & 66.5 & \textbf{78.0} & 50.6 & \textbf{80.9} & \textbf{89.1} & 5.9 & 18.2 & 27.1 & 9.5 & 24.7 & 34.8  \\
SpeechCLIP-C \cite{speechclip2022}
                    & 14.7 & 41.2 & 55.1 & 21.8 & 52.0 & 67.7 & 6.4 & 20.7 & 31.0 & 9.6 & 27.7 & 39.7 & - & - & - & - & - & -  \\
ResDAVEnet \cite{resdave}
                     & - & - & - & - & - & - & 17.3 & 41.9 & 55.0 & 22.0 & 50.6 & 65.2 & - & - & - & - & - & - \\
MILAN \cite{talkdontwrite2021}

                     & 33.2 & 62.7 & 73.9 & 49.6 & 79.2 & 87.5 & - & - & - & - & - & -  & - & - & - & - & - & -  \\
FaST-VGS-co \cite{Fast-VGS}
                    & 26.6 & 56.4 & 68.8 & 36.2 & 66.1 & 76.5 &  31.8 & 62.5 & 75.0 & 42.5 & 73.7 & 84.9 & - & - & - & - & - & -   \\
FaST-VGS-ctf \cite{Fast-VGS}
                    &  29.3 & 58.6 & 71.0 & 37.9 & 68.5 & 79.9 & 35.9 & 66.3 & 77.9 & 48.8 & 78.2 & 87.0 & - & - & - & - & - & -  \\
\midrule
\textbf{Our Experiments}\\  
\midrule
PIP-Only BrewCLIP(zero-Shot) 
                    & 58.8 & 83.9 & 90.5 & 77.4 & 93.4 & 97.0  & 33.6 & 57.7 & 68.0 & 53.9 & 77.8 & 85.4 & 16.3 & 33.0 & 42.1 & 23.6 & 41.5 & 50.8  \\

PIP-Only BrewCLIP (Prompted) 
                    & \textbf{68.5} & \textbf{89.9} & 94.5 & \textbf{80.8} & 95.8 & 98.6  & \textbf{42.2} & 67.0 & 76.3 & 53.6 & 77.9 & 85.7 & 39.1 & 63.2 & 72.7 & 40.0 & 63.1 & 71.7 
                    \\
BrewCLIP
                     & 66.5 & 89.1 & \textbf{94.6} & 79.3 & \textbf{95.9} & \textbf{98.6}  & 40.7 & \textbf{67.1} & \textbf{77.5} & \textbf{54.2} & 79.6 & 87.8 & \textbf{40.3} & \textbf{68.1} & \textbf{78.7} & \textbf{42.3} & \textbf{67.6} & \textbf{77.6} 
                    \\
\bottomrule
\end{tabular}
\end{adjustbox}
\caption{\textbf{Comparison between our proposed models and previous SotA methods on the audio-image bidirectional retrieval task.}}
\label{tab:result_SoTA}
\end{table*}

\begin{table*}[htb!]
\vspace{-6pt}
\begin{adjustbox}{max width=0.95\textwidth,center}
\begin{tabular}{*{15}{c}}
\toprule                  

                            & \multicolumn{6}{c}{SpokenCOCO}
                            & \multicolumn{6}{c}{LN-COCO}
                            \\
                            \cmidrule(lr){3-6}\cmidrule(lr){9-12}
                            & \multicolumn{3}{c}{Speech2Image} 
                            & \multicolumn{3}{c}{Image2Speech}
                            & \multicolumn{3}{c}{Speech2Image}
                            & \multicolumn{3}{c}{Image2Speech}

                            \\
 \cmidrule(lr){2-4}\cmidrule(lr){5-7}\cmidrule(lr){8-10}\cmidrule(lr){11-13}
Method & R1  & R5  & R10 & R1  & R5  & R10 & R1  & R5  & R10 & R1  & R5  & R10 \\

\midrule
BrewCLIP
                     & 40.7 & \textbf{67.1} & \textbf{77.5} & \textbf{54.2} & \textbf{79.6} & \textbf{87.8} & \textbf{40.3} & \textbf{68.1} & \textbf{78.7} & \textbf{42.3} & \textbf{67.6} & \textbf{77.6} 
                    \\
E2E-only BrewCLIP 
                    & 22.8 & 50.6 & 64.4 & 35.2 & 65.8 & 77.8 & 24.0 & 51.6 & 65.0 & 28.4 & 53.1 & 64.9   \\                
PIP-only BrewCLIP (prompted) & \textbf{42.2} & 67.0 & 76.3 & 53.6 & 77.9 & 85.7 &  39.1 & 63.2 & 72.7 & 40.0 & 63.1 & 71.7 
                    \\ 
\midrule      
\rowcolor{mygray}
PIP-Only BrewCLIP (Prompted) on GT  
                    & 44.9 & 69.8 & 78.9 & 55.8 & 79.2 & 86.9 & 48.1 & 72.8 & 81.3 & 47.5 & 71.3 & 79.7 \\ 

\bottomrule
\end{tabular}
\end{adjustbox}
\caption{\textbf{Comparison between different variations of our models (pipeline-only, E2E-only, and full model) on the audio-image retrieval task for SpokenCOCO, and LN-COCO.}}
\label{tab:result_modeldif}
\vspace{-6pt} 
\end{table*}
\begin{figure*}[!]
\centering
{\includegraphics[width=15.5cm]{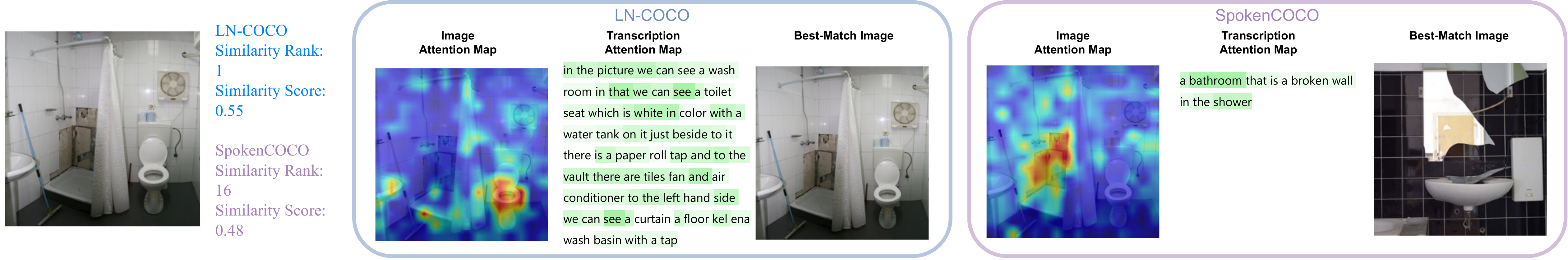}}
\caption{ \textbf{Qualitative analysis of sample difference between LN-COCO and SpokenCOCO and their impacts on model performance.} Audio descriptions from Flick8K and SpokenCOCO typically consist of concise, pre-planned single sentences. Such scripted descriptions pose for the recall task as they might match with multiple similar images. The LN-COCO samples contains much more detailed information to help the model localize the exact match of the image. 
In the Audio2Image matching task, \textbf{Similarity Rank} refers the rank of the current image relative to all provided images,determined on Cosine similarity score and the \textbf{Best-Match Image} is the one that has the highest similarity score to the given audio. \textbf{(More examples in Appendix)}}
\label{fig:data_diff}
\vspace{-5mm}
\end{figure*}
\noindent \textbf{RAVDESS}
This dataset comprises 1440 audio files recorded by 24 professional actors and labeled with eight different emotional expressions: neutral, calm, happy, sad, angry, fearful, disgust, and surprised. For our investigation, we focus on four regular emotions, namely neutral, happy, sad, and angry, consistent with previous studies. This dataset contain speech clips with the same textual content but different emotional tone. This aspect makes the dataset ideal for testing whether our model can successfully capture non-textual information contained in spoken communication. 
\section{Results and Discussion}
 \begin{figure*}[t]
\vspace{-8ex}
\centering
{\includegraphics[width=15.5cm]{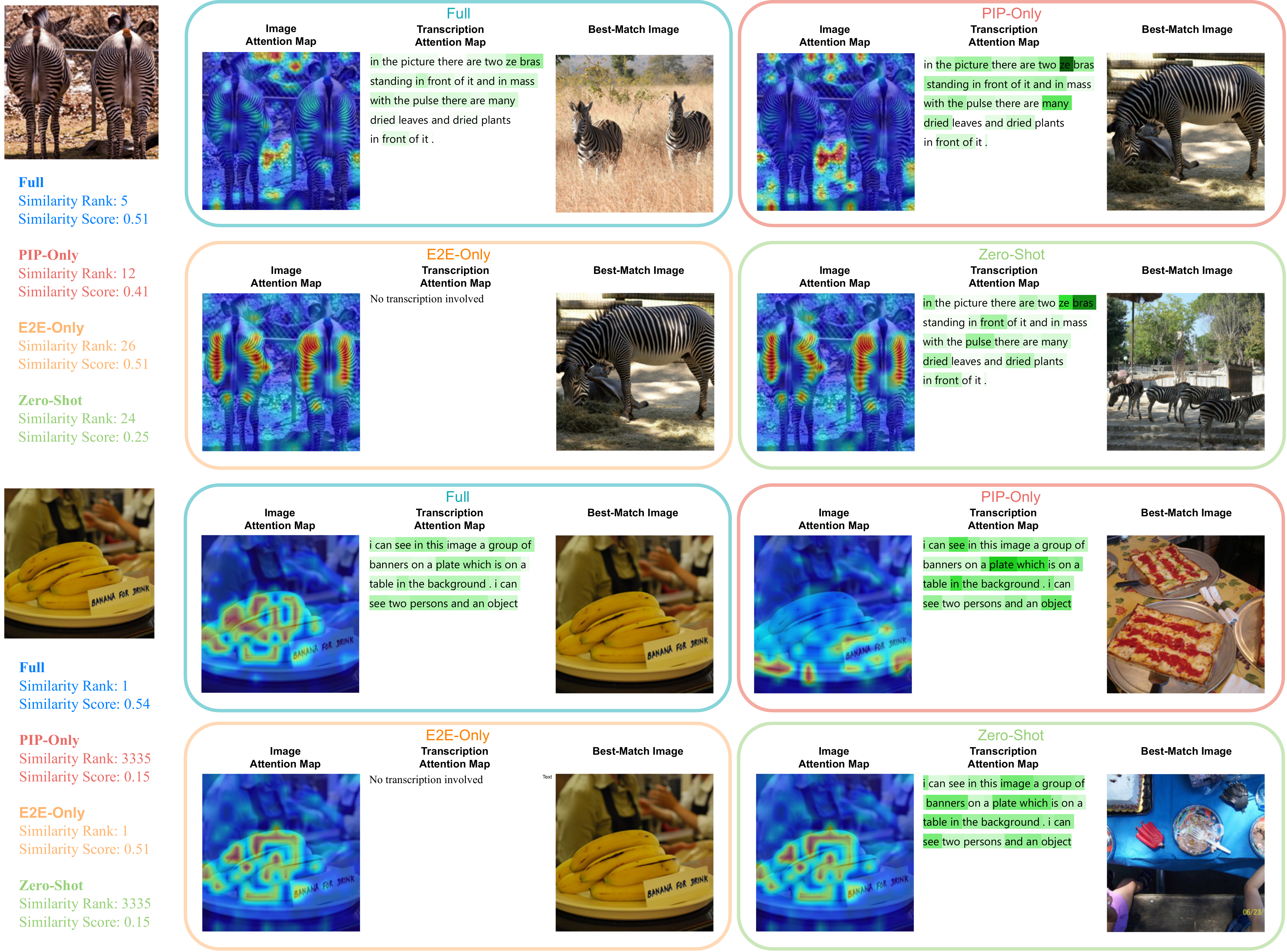}}
\caption{ \textbf{Qualitative analysis of different variations of models on audio to image retrieval.} The top case shows our full model can attend to non-major content, making optimal use of the rich information in the LN-COCO samples, whereas E2E model and Zero-Shot pipeline model focus solely on the major components. The bottom case shows our full model can still successfully retrieve the correct image, even the transcription has a major mistake, underscoring the necessity of the inclusion of the E2E channel.\textbf{(More examples in Appendix)}}
\label{fig:model_diff_qual}
\vspace{-5mm}
\end{figure*}
\textbf{Comparison with SotA}
As indicated in Table. \ref{tab:result_SoTA}, the pipeline-only model, even under the zero-shot setting, has already surpassed existing works that were trained or fine-tuned by target dataset. Moreover, our final model achieves even higher scores.
\\
To the best of our knowledge, no previous studies have evaluated audio-image retrieval on the LN-COCO dataset. Therefore, as a baseline, we implemented one of the most recent and well-performing E2E models, the parallel version of SpeechCLIP (SpeechCLIP-P), which has achieved SotA results on Flickr8k and SpokenCOCO dataset to test its performance on an unscripted dataset. As Shown in Table. \ref{tab:result_SoTA}, SpeechCLIP-P's performance on LN-COCO is much lower than BrewCLIP's. 
\\
\textbf{Hypothesis Validation}
To test our hypothesis that non-textual information is useful for the downstream task, we compare three variations of our model: the full model, pipeline-only model and E2E-only model. The E2E module is similar to the SotA SpeechCLIP-P but with the Whisper encoder replacing the Hubert model. We experiment on both the scripted SpokenCOCO and unscripted LN-COCO, which share the same source image set, enabling a fair comparison. As depicted in Table. \ref{tab:result_modeldif}, the pipeline-only model generally outperforms the E2E-only model, especially when a smaller number of guesses are allowed in the recall task. This observation is consistent with existing literature and indicates that transcribed text as an intermediate output, combined with a robust text encoder, better enables the learning of subtle and detailed information present in the paired utterances and images, benefiting retrieval tasks (qualitative analysis shown in Fig. \ref{fig:data_diff}). This difference becomes particularly evident when experimenting on the unscripted LN-COCO dataset, where the performance of the E2E model significantly lags behind the pipeline model.
 For the scripted SpokenCOCO, the performance of the pipeline-only model does not show a significant difference compared to the full BrewCLIP model. This is as expected since transcribing is easier for scripted utterances, and they might not contain as much non-textual information. On the other hand, we observe a significant improvement after adding the E2E module to the pipeline-only model when evaluating our full model on the unscripted LN-COCO dataset. To further investigate whether this improvement comes from the E2E channel compensating for mistakes made by the ASR model or from the non-textual information, we directly feed the exact ground truth captions into our pipeline-only model. As shown in Table. \ref{tab:result_modeldif}, the improvement is more likely attributed to the former as feeding ground truth captions into our pipeline-only model can even surpass our full model. To this end, we can conclude that the E2E channel can act as a remedy if the transcription process fails, but whether our model can effectively extract non-textual information remains inconclusive. \\
\textbf{Multimodal Dataset Aware of Non-Texutal Content}
To further test the capabilities of our proposed model, we recognize the need for a new dataset that emphasizes extra information beyond purely textual content. This dataset should focus on capturing elements such as tone, emotion, and other non-semantic aspects present in the audio, as they are likely to play a crucial role in determining the correct pairing of audio and image. 

\begin {table}[htb!]
\centering

\begin{center}

\resizebox{0.45\textwidth}{!}{%
\begin{tabular}{*{3}{c}}
\toprule            
Method& & Accuracy\\
\midrule
BrewCLIP  &Linear Prob & 56 \\ 
&Finetuned & 78 \\
\midrule
PIP-Only BrewCLIP &Linear Prob & 33 \\ 
&Finetuned & 33 \\
\bottomrule
\end{tabular}
}
\caption{\textbf{SER Result on RAVDESS.} Finetuned means only prompts and final transformer encoder layer in the audio channel are updated.}
\label{tab:SER}
\end{center}
\vspace{-6pt} 
\vspace{-4mm}
\end{table}
\noindent \textbf{Speech Emotion Recognition} We fail to find such a dataset. As an alternative, we conduct a simple Speech Emotion Recognition (SER) experiment only testing our trained audio encoder. For this experiment, we train an additional linear classification head to test whether our model can accurately recognize emotions in the utterances. As shown in Table. \ref{tab:SER}, As anticipated, the pipeline-only model does not work as the textual content is the same for each utterance. Nonetheless, we observe some effects by applying a simple linear classifier head to our frozen audio encoder with even higher performance achieved through finetuning. These results robustly support our hypothesis that our proposed model can successfully capture critical non-textual information (primarily mood changes in this case) that exists in the representation learned from the E2E module. This non-textual information plays a significant role in constructing a holistic audio representation. \\
\textbf{Effect of Prompting} 
A considerable portion of the model's mistakes are related to successfully transcribing the text but failing in a manner typical of CLIP, such as paying attention to common objects in the image rather than focusing on more specific elements. 
This issue was more pronounced in the unscripted LN-COCO dataset, where the abundance of details often confuses the CLIP model, resulting in the poor performance of  Zero-Shot pipeline model.  The shared prompting technique is effective in ameliorating the issues by establishing a connection between different modalities and has shown to enable the model to comprehend long and detailed LN-COCO utterances as shown in Fig. \ref{fig:model_diff_qual}.



\begin {table}[htb!]
\centering
\begin{center}
\resizebox{0.45\textwidth}{!}{%
\begin{tabular}{*{9}{c}}
\toprule                  
                            & \multicolumn{3}{c}{Speech2Image} 
                            & \multicolumn{3}{c}{Image2Speech}
                            
                            \\
\cmidrule(lr){2-4}\cmidrule(lr){5-7}
Method & R1  & R5  & R10 & R1  & R5  & R10 \\
\midrule
SpokenCOCO\\  
\midrule
ASR transcription & 33.6 & 57.7 & 68.0 & 53.9 & 77.8 & 85.4\\ 
Ground truth & 36.1 & 60.8 & 71.1 & 56.2 & 78.8 & 86.9  \\
\midrule
LN-COCO\\  
\midrule
ASR transcription & 16.3 & 33.0 & 42.0 & 23.6 & 41.5 & 50.8\\
Ground truth & 20.8 & 40.9 & 51.4 & 28.6 & 50.8 & 60.7\\
\bottomrule
\end{tabular}
}
\caption{\textbf{Comparison of using ASR output vs ground truth in our Zero-Shot models.}}
\label{tab:result_wer}
\end{center}
\end{table}
\noindent \textbf{Impact of Transcription Quality}
We evaluated the model's performance by comparing using both generated transcriptions and ground truth captions. As shown in Table. \ref{tab:result_wer}, for the SpokenCOCO dataset, the model perform slightly better on the ground truth text compared to the transcriptions. However, for the LN-COCO dataset, the performance gap is significantly larger due to a higher error rate in the transcriptions. We also analyze the per-sample Word Error Rate (WER) in relation to its recall accuracy and fits it into a logistic regression model. The results clearly demonstrates a clear pattern where the per-sample recall is inversely related to the WER, and confirms that ASR errors can have a substantial impact on the model's performance, further highlighting the need of an additional audio channel which has been proved to have the capability to correct these mistakes. 

\noindent \textbf{Cross Dataset Evaluation}
To test the generalizability of our fine-tuned model on different dataset, we evaluate all three full BrewCLIP models in a cross-dataset way. Table.
\ref{tab:cross_dataset} shows the models trained with Flickr8K and SpokenCOCO adapt each other very well. However, the models trained with LN-COCO fail to adapt to Flickr8K and even SpokenCOCO which shares the same image set. Non-finetuned CLIP text encoders struggle with processing very long sentences and our prompting design successfully addresses this limitation but, unfortunately, compromising its original capability of processing short sentences.\\ 
\begin {table}[htb!]
\centering
\begin{center}
\vspace{-20pt} 
\resizebox{0.45\textwidth}{!}{%
\begin{tabular}{*{10}{c}}
\toprule       \textbf{Model-Dataset}           
                            & \multicolumn{1}{c}{LN-COCO} 
                            & \multicolumn{1}{c}{SpokenCOCO}
                            & \multicolumn{1}{c}{Flickr8k}
                            \\
\midrule
LN-COCO & 68.1/78.6 & 38.8/58.2 & 53.6/76.1\\ 
\midrule
SpokenCOCO & 47.3/41.9 & 67.1/79.6 & 87.8/93.8\\ 
\midrule
Flickr8K & 38.9/40.4	& 61.3/75.3	& 89.1/95.9\\ 
\bottomrule
\end{tabular}
}
\caption{\textbf{Cross-dataset Evaluation.} Column denotes the dataset that the model is trained with and row denotes the dataset evaluated on.Value entry follows the format: \textbf{R5 @ Speech2Image / R5 @ Image2Speech}}
\label{tab:cross_dataset}
\end{center}
\vspace{-6pt} 
\vspace{-4mm}
\end{table}

\begin{figure}[!]
\centering
 \includegraphics[scale=0.22]{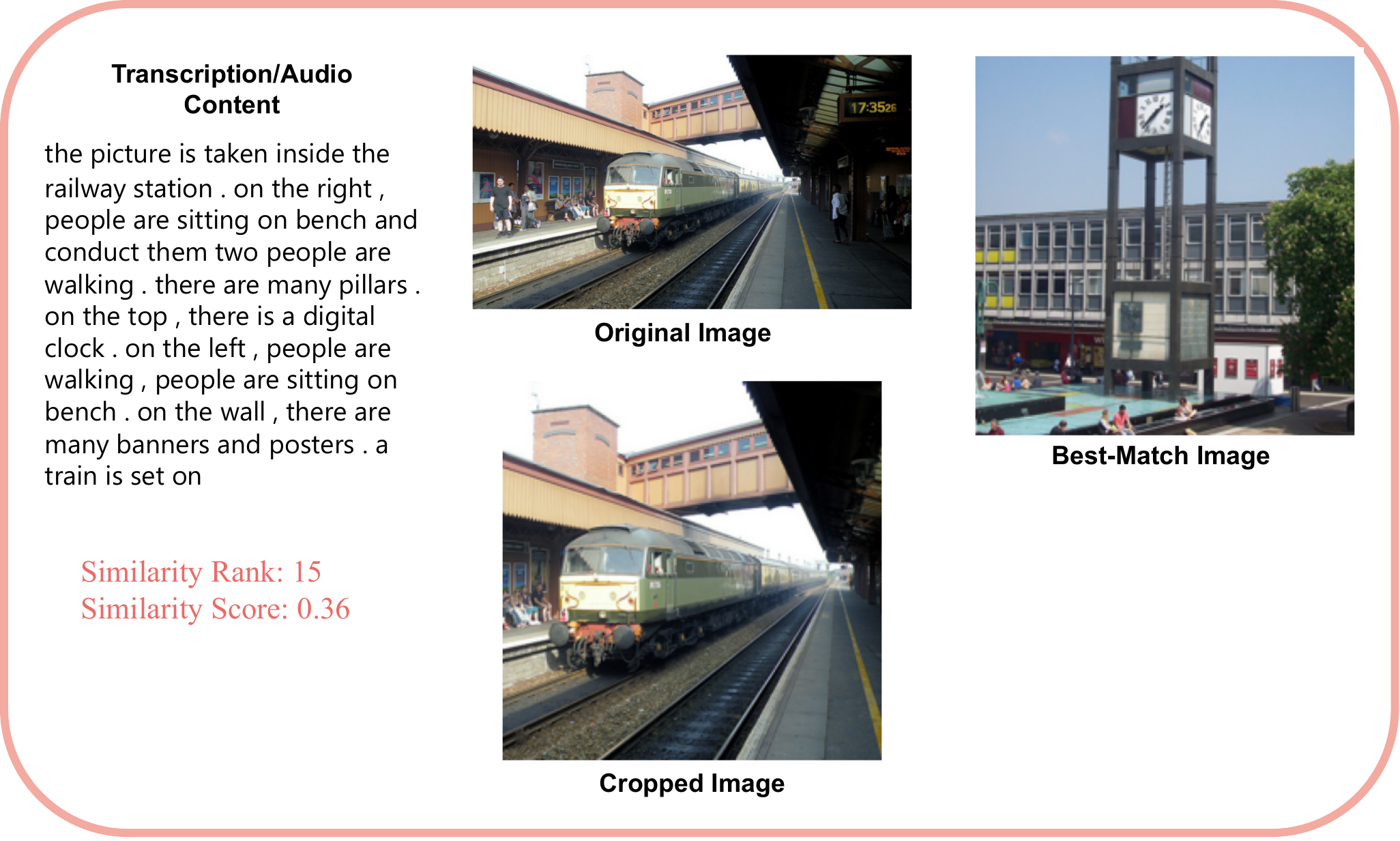}
\caption{\textbf{Failed case due to the Image Center Crop transform.} The original image features a clock, a detail referenced in the spoken expression; however, the clock is cropped in the image.}
\label{fig:crop_error}
\end{figure}
\newpage
\noindent \textbf{Pitfalls of CLIP Image Transform}
CLIP incorporates the center crop strategy as part of its image pre-processing. Many errors made by the model is attributed to the absence of information from the region cropped by the transformation function, leading to the discrepancy between the visual content and its associated audio as shown in Fig. \ref{fig:crop_error}.


\section{Conclusion}
In this work, we argue that current pipeline-based approaches for audio-image retrieval, while achieving remarkable performance, inevitably lose critical audio-specific information during the transcribing process. To address this issue, we propose a novel dual-channel architecture that effectively learns a compact joint representation between speech and image. Our study not only establishes a robust baseline for zero-shot pipeline models but also pushes the boundaries of the current state-of-the-art. Furthermore, we run comprehensive experiments evaluating various types of model on different types of dataset, elucidating the importance of employing a dual-channel model to build a more comprehensive image-audio joint representation.    



\section{Limitation}
Our primary limitation lies in the inability to directly demonstrate that our audio-image joint embedding can capture the shared non-textual connection between image-audio pairs, due to the absence of a suitable dataset. Also, our model faces constraints imposed by the pre-trained models used. The CLIP text encoder can only process at most 77 tokens while some LN-COCO samples extend up to 2 minutes, exceeding the 77-token limit when transcribed. This results in the truncation of potentially important information. Also, CLIP image transformation is also a potential avenue for enhancement as discussed above. 


\bibliography{reference}
\appendix
\section{Detailed Implementation}
For the ASR model, we apply base (74M) model of Whisper model. Since the Whisper model is frozen entirely, we save the transcriptions as the intermediate output to avoid the need for inference every epoch.\\
As for text-image model, we opt for ViT-L/14 (427M) as our pre-trained CLIP model, since the transformer-based architecture of ViT-L/14 facilitates the use of prompting for fine-tuning.

Following the standard of SpeechCLIP, all models are trained with Adam optimizer with a weight decay of $10^{-6}$, batch size of 32. The maximum training step is set to 100k steps in total. The learning rate linearly increases to $10^{-4}$ in the first 5k steps and linearly decreases to $10^{-8}$. The hyperparameter to control the contribution of the inner and outer losses, $\alpha$ is set to be 0.1 and prompt length, $m$ is set to be 4.  All experiments can be conducted on three NVIDIA RTX A4000 GPUs \\
The attention map is generated by computing the gradient-updating direction of the penultimate layer of each module. \\
\textbf{Corruption of Audio Samples}
Some audio files either contain portion of corrupted signal, which actually causes the ASR model to crash and therefore, not able to generate transcriptions successfully, especially in the Loconarr-COCO dataset. There are 21 audio samples, out of 8573 samples, that cause the ASR model to crash in the LocoNarr-COCO validation dataset, and therefore we simply drop them. 

 \begin{figure*}[]
\vspace{-5mm}
\centering
{\includegraphics[width=15.5cm]{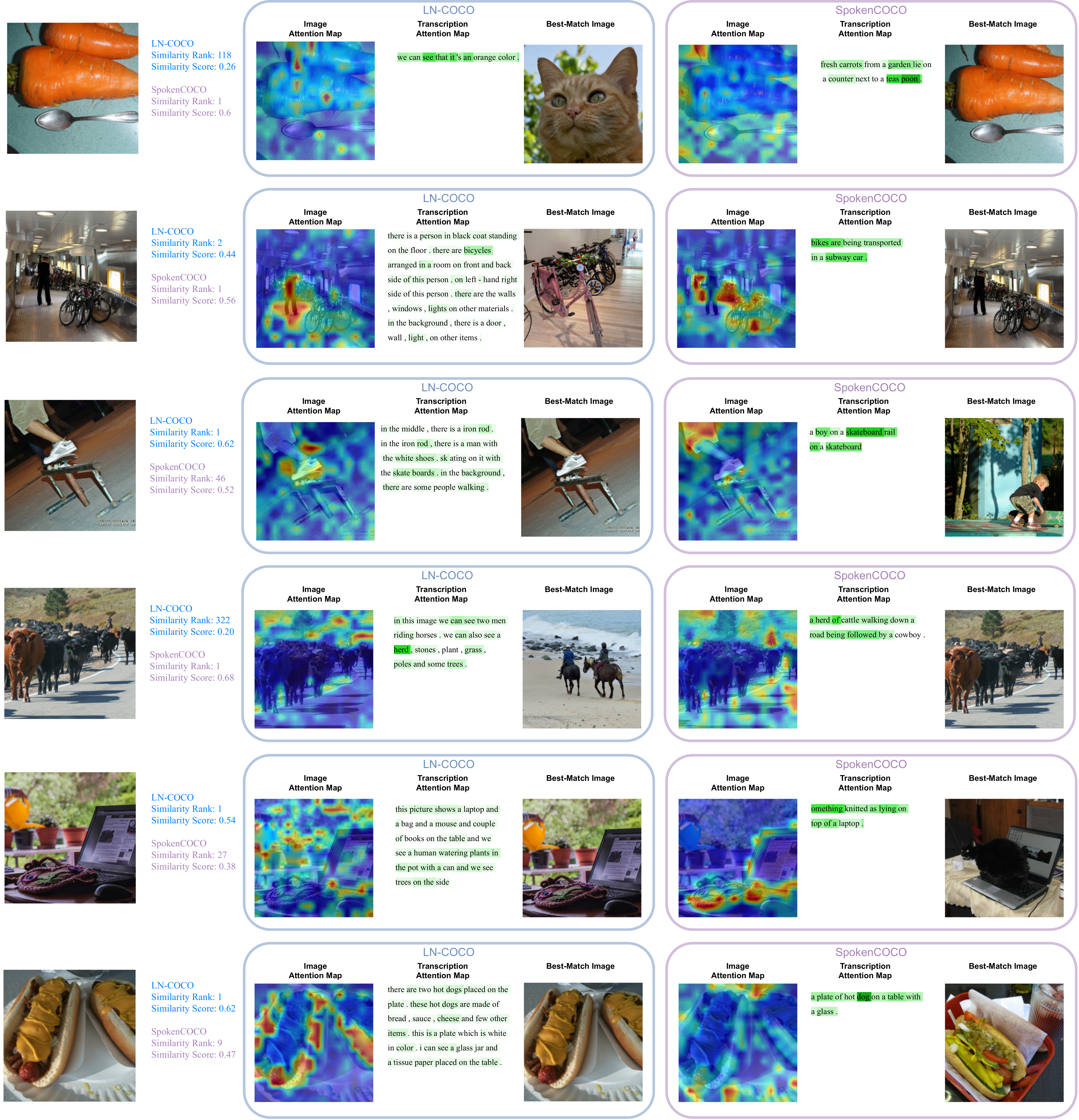}}
\caption{ \textbf{Qualitative analysis of between LN-COCO and SpokenCOCO and their impacts
on model performance}}
\vspace{-5mm}
\end{figure*}

 \begin{figure*}[htbp]
\vspace{-5mm}
\centering
{\includegraphics[width=15.5cm]{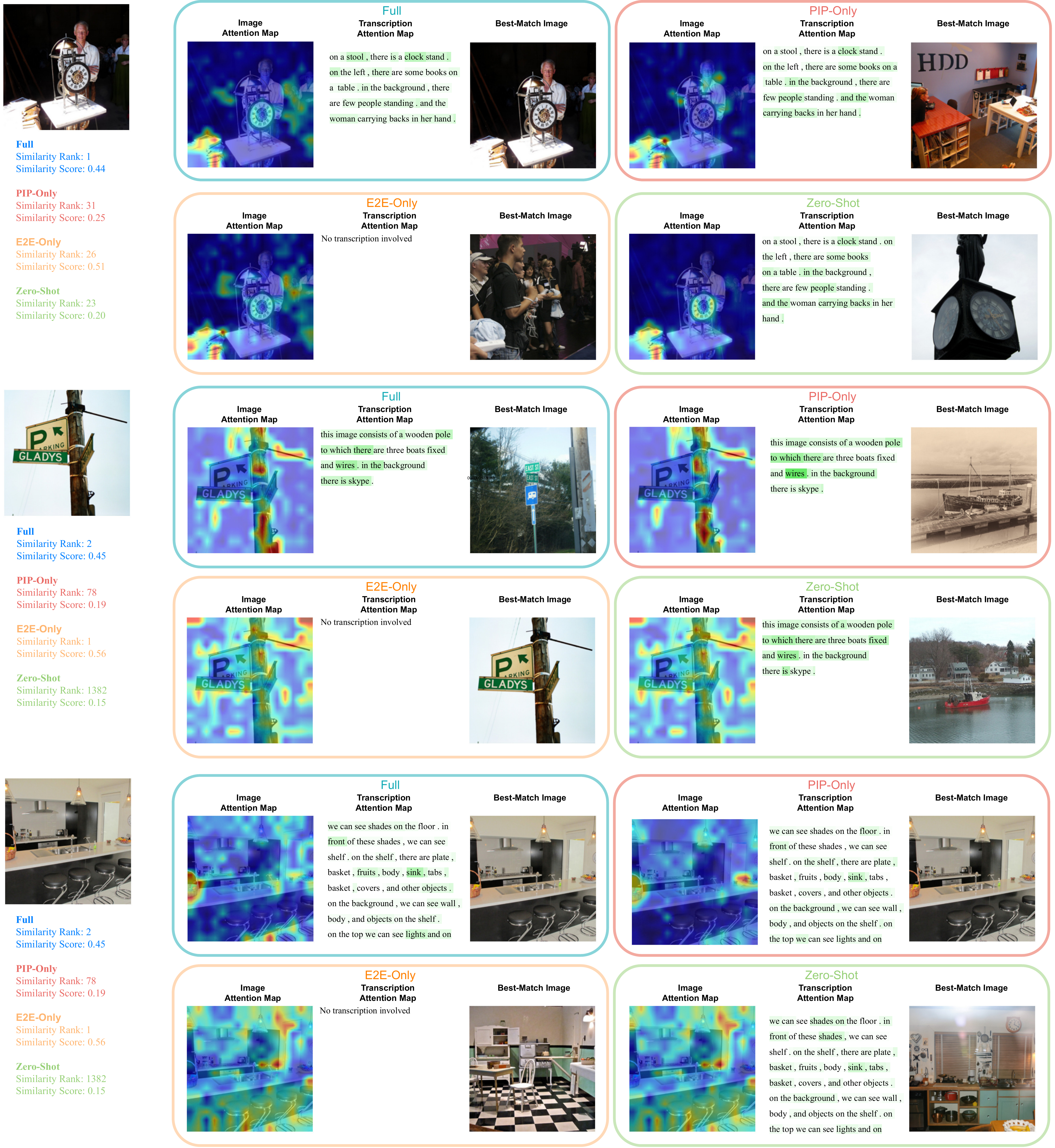}}
\caption{ \textbf{Qualitative analysis of different variations of models on audio to image retrieval.}}
\label{fig:model_diff_app}
\vspace{-5mm}
\end{figure*}
 \begin{figure*}[htbp]
\vspace{-5mm}
\centering
{\includegraphics[width=15.5cm]{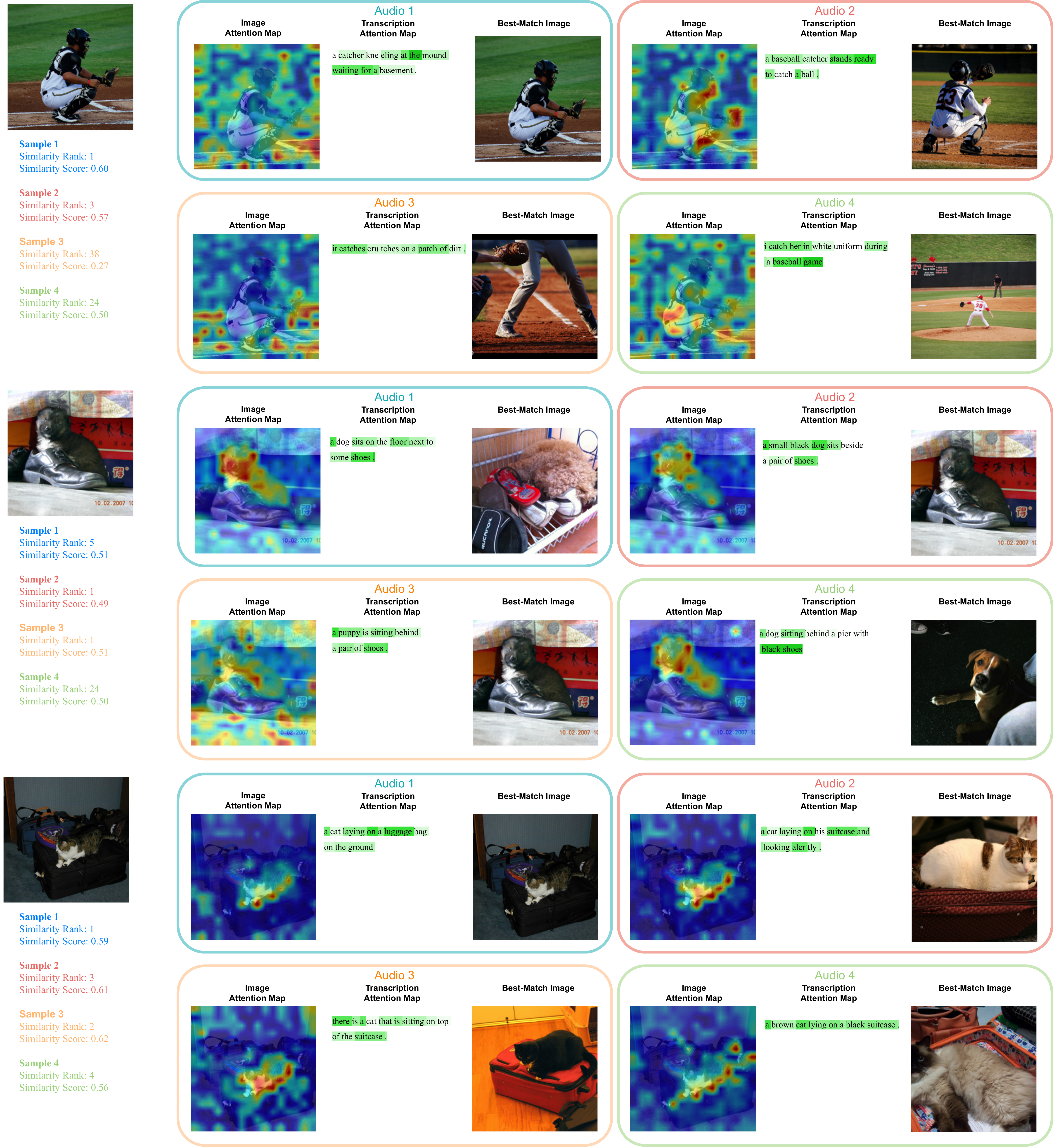}}

\caption{ \textbf{Qualitative analysis of different audio clips from SpokenCOCO and their impact on audio to image retrieval}}
\label{fig:sample_diff_app}
\vspace{-5mm}
\end{figure*}
 \begin{figure*}[htbp]
\vspace{-5mm}
\centering
{\includegraphics[width=15.5cm]{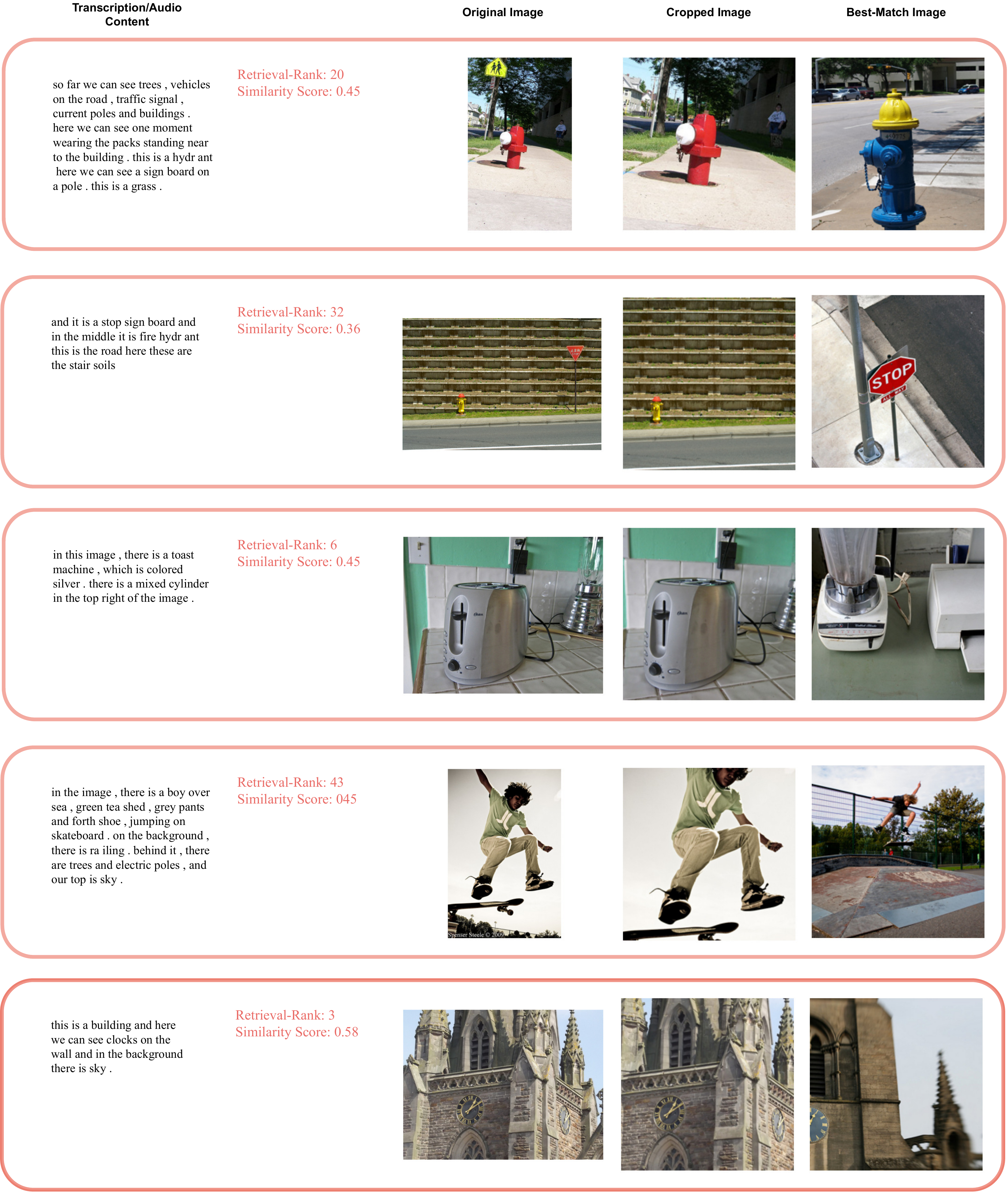}}
\caption{ \textbf{Failed case due to the Image Center Crop
Transform}}
\label{fig:crop_error_app}
\vspace{-5mm}
\end{figure*}

\end{document}